\documentclass[letterpaper,twocolumn,10pt]{article}
\usepackage{template/usenix-2020-09}

\usepackage{tikz}
\usepackage{amsmath}
\usepackage{amsthm}

\usepackage{filecontents}

\usepackage{amsfonts}
\usepackage{url}
\usepackage{graphicx}
\usepackage{booktabs}
\usepackage{multirow}
\usepackage{threeparttable}
\usepackage{amssymb}
\usepackage{mathtools}
\usepackage{subcaption}
\usepackage{svg}
\usepackage{comment}
\usepackage{algorithm}
\usepackage{algpseudocode}
\usepackage{amsmath}

\usepackage{listings}

\newcommand{\ID}{\mathit{ID}}
\newcommand{\esk}{\mathit{esk}}
\newcommand{\lsk}{\mathit{lsk}}
\newcommand{\pk}{\mathit{pk}}
\newcommand{\K}{\mathit{K}}
\newcommand{\SK}{\mathit{SK}}

\newcommand{\senc}{\mathrm{senc}}

\newcommand{\aenc}{\mathrm{aenc}}

\newcommand{\sign}{\mathrm{sign}}

\newcommand{\hash}{\mathrm{hash}}

\newcommand{\sendIR}{\mathrm{sendIR}}
\newcommand{\sendRI}{\mathrm{sendRI}}

\newcommand{\AN}{\mathbf{AN}}
\newcommand{\FN}{\mathbf{FN}}
\newcommand{\BN}{\mathbf{BN}}
\newcommand{\Pub}{\mathbf{Pub}}

\theoremstyle{remark}

\newtheorem*{example*}{Example}

\begin{document}

\date{}

\title{\Large \bf A Security Verification Framework of Cryptographic Protocols \\Using Machine Learning}

\author{
{\rm Kentaro Ohno}\\
NTT Computer \& Data Science Laboratories
\and
{\rm Misato Nakabayashi}\\
NTT Social Informatics Laboratories
} 

\maketitle

\begin{abstract} 

We propose a security verification framework for cryptographic protocols using machine learning.
In recent years, as cryptographic protocols have become more complex, research on automatic verification techniques has been focused on.
The main technique is formal verification. However, the formal verification has two problems: it requires a large amount of computational time and does not guarantee decidability.
We propose a method that allows security verification with computational time on the order of linear with respect to the size of the protocol using machine learning.
In training machine learning models for security verification of cryptographic protocols, a sufficient amount of data, i.e., a set of protocol data with security labels, is difficult to collect from academic papers and other sources.
To overcome this issue, we propose a way to create arbitrarily large datasets by automatically generating random protocols and assigning security labels to them using formal verification tools.
Furthermore, to exploit structural features of protocols,
we construct a neural network that processes a protocol along its series and tree structures.
We evaluate the proposed method by applying it to verification of practical cryptographic protocols.

\end{abstract}

\section{Introduction} 
Today, cryptographic protocols are used in a variety of important situations and are indispensable technologies.
For example, TLS 1.3 is used for confidentiality, tamper detection, and authentication of communication partners on the Internet~\cite{TLSstd}.
Vulnerabilities in cryptographic protocols can compromise communications over the Internet and tamper with electronic transactions. The impact is serious, and therefore the security of cryptographic protocols is important.
However, the design of cryptographic protocols is generally complex and error-prone.

To design complex and correct cryptographic protocols, designers often use computer aids 
such as automated verification tools.
As we input the protocol specification and security requirements, an ideal automated verification tool instantly outputs whether the protocol satisfies the security requirements or not.
A typical example of such an automatic verification tool is a formal verification tool based on model checking.
Formal verification is a technique that describes a target using a formal language and verifies whether or not the target satisfies certain requirements by using mathematical techniques.
In particular, formal verification tools based on model checking provide exhaustive verification by thoroughly enumerating possible states of the target and verifying all possible paths to the target.
There are many formal verification tools for cryptographic protocols; ProVerif~\cite{blanchet2013automatic,proverifweb} and Tamarin prover~\cite{tamarinweb,meier2013tamarin} are well-known examples.
These tools are used in the design of widely used cryptographic protocols and contribute to the design of secure protocols.
For example, formal verification tools were used in the standardization process of TLS 1.3 and 5G authentication protocols, and many vulnerabilities were found by these tools~\cite{8970242,bhargavan2022symbolic,bhargavan2017verified,delignat2017implementing,blanchet2018composition,cremers2017comprehensive,cremers2016automated,delignat2021security,basin2018formal,cremers2019component}.
As shown in these examples, the strength of formal verification tools is that they can find vulnerabilities that are difficult for the human eye to detect.

The problem with formal verification is the time required to perform exhaustive verification.
In addition, verification may never be completed because the verification tools do not have decidability~\cite{durgin2004multiset,tiplea2005decidability}.
For example, the TLS verification by Bhargavan et al.~\cite{bhargavan2022symbolic} took up to 35 hours.
Therefore, the verifier should devise ways to formalize the protocol and use the tools to complete the verification within the effective time.
This requires high-level expertise.
This is one reason that many excellent formal verification tools for cryptographic protocols exist but have not received enough attention in the industry.

\begin{figure*}[t]
  \begin{minipage}[b]{0.48\linewidth}
    \centering
    \includegraphics[height=6.5cm]{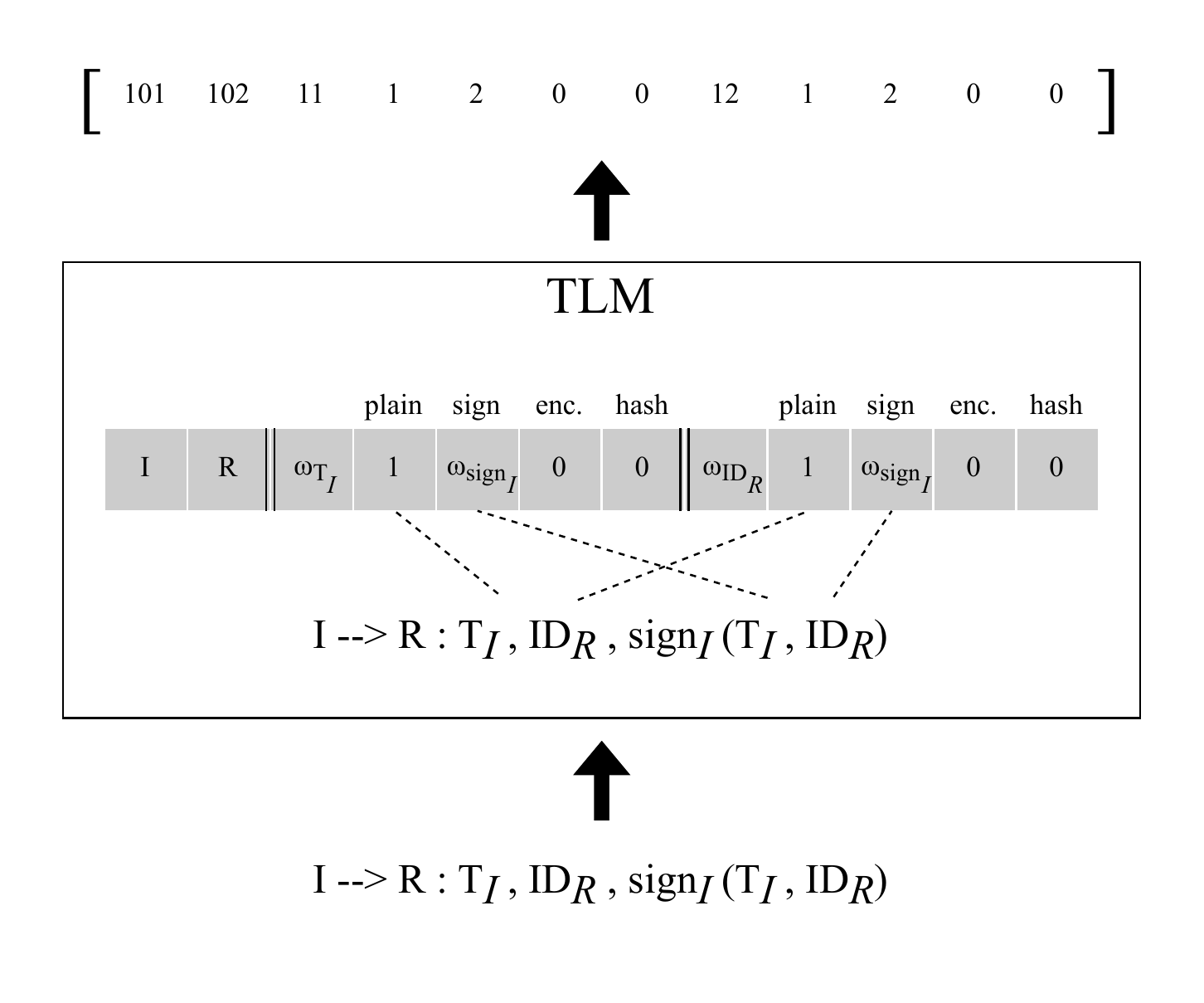}
    \subcaption{
    An example of protocol conversion by the method TLM
    ~\cite{ma2020machine}.
    }
    \label{fig:protocol_feature_example}
  \end{minipage}\hspace{1cm}
  \begin{minipage}[b]{0.45\linewidth}
    \centering
    \includegraphics[height=6.5cm]{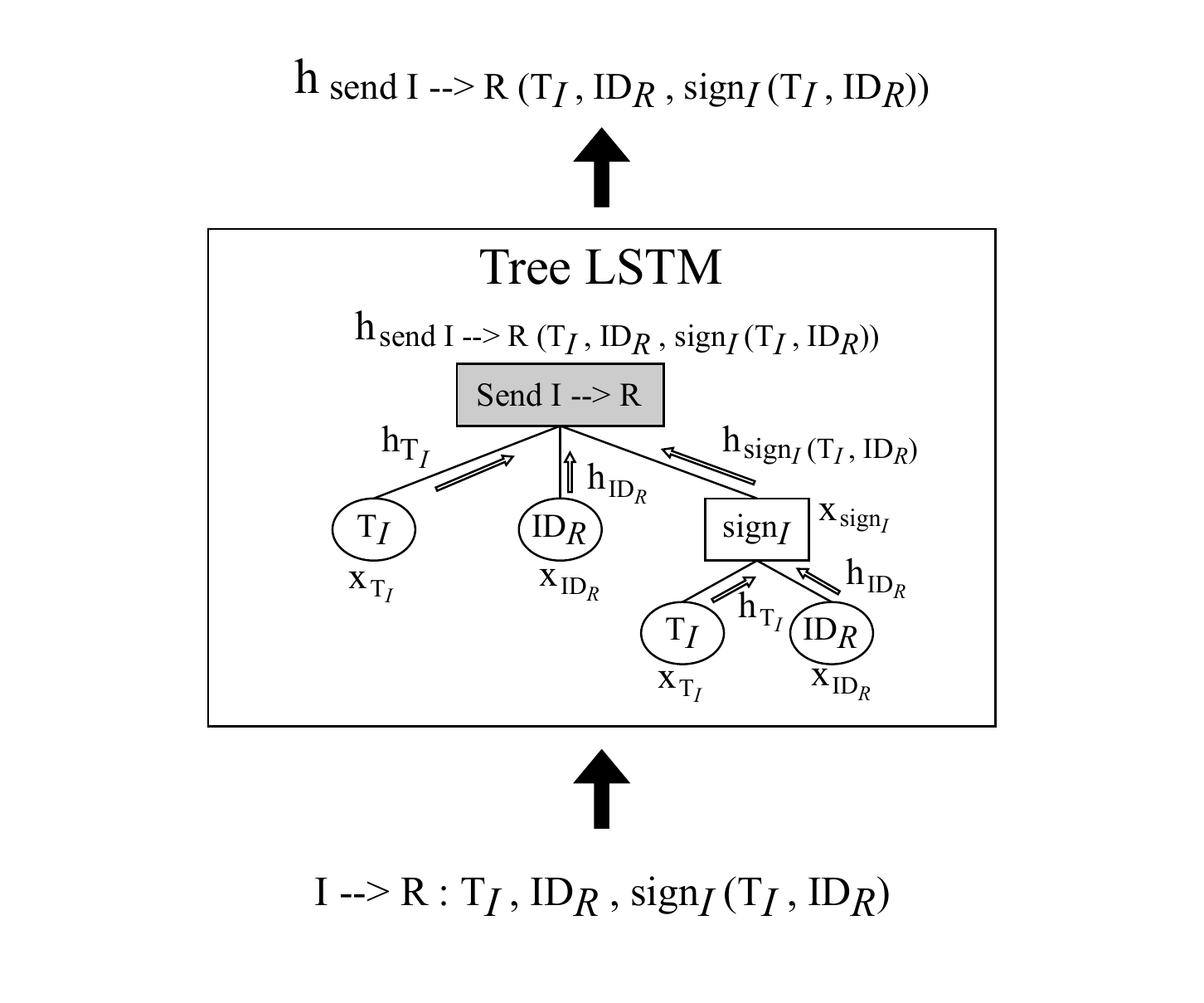}
    \subcaption{
    An example of protocol conversion by our proposed method.
    }
    \label{fig:protocol_feature_ours}
  \end{minipage}
  \caption{
  Comparison of protocol conversion between the previous method~\cite{ma2020machine} and our proposed method. The previous method loses information on the hierarchical structure such as the order of application of operations such as signing and encryption during the conversion (see Section~\ref{subsec:const_ml}). In contrast, the proposed method enables the conversion that reflects the order of application of operations by using a neural network that incorporates a tree structure.}
\end{figure*}

To perform verification in practical time, we focus on machine learning techniques, in particular deep learning, as a verification tool.
In general, 
a machine learning model makes inferences on a single input decisively in a short time.
Thus, the verification of the security of cryptographic protocols is expected to always terminate in a short period of time.
We consider using deep learning, which is expected to be highly accurate among machine learning methods.
Deep learning has achieved great successes in a wide range of areas such as image classification~\cite{he2016deep}, image generation~\cite{ho2020denoising}, and language processing~\cite{brown2020language}, particularly where abundant training data are available.
The goal of this paper is to construct a verifier for cryptographic protocols using machine learning that is practically fast, decidable, and accurate.

To achieve this, large amounts of training data and an appropriate model are desirable~\cite{he2016deep,brown2020language}.
At present, such a verifier using machine learning has not been realized.
One reason is the difficulty of collecting training data.
The training data 
consists of pairs of cryptographic protocols and their security label.
This information is usually obtained from papers, textbooks, and standardization documents.
They are few in number, and there are not enough pairs of cryptographic protocols and their security labels for training.
Therefore, a sufficient amount of training data is difficult to collect.
There are two previous studies using machine learning to verify the security of cryptographic protocols~\cite{ma2020machine,zahednejad2020novel}, but the number of training data in those studies is only 500 to 1000, which is insufficient.
Another issue is construction of machine learning models suitable for processing cryptographic protocols.
Since cryptographic protocols can easily violate the security requirements if the protocols are slightly changed~\cite{boyd2019protocols},
detailed information of protocols including hierarchical structures of messages in the protocols should be encoded to a vector by a machine learning model.
However, this makes it difficult to construct a model that can appropriately handle cryptographic protocols.
Previous studies have proposed methods for converting protocols to vectors of fixed length, but there is no one-to-one correspondence between protocols and converted vectors, i.e., multiple different protocols are converted to the same vector (Figure~\ref{fig:protocol_feature_example}).
Such conversion causes the loss of structural information of the cryptographic protocols, which prevents the achievement of high verification accuracy.
Therefore, for accurate protocol verification, machine learning models need to be generated that reflect information on hierarchical structures that are ignored by existing methods.
The construction of a model that adapts to the data structure of cryptographic protocols is expected to improve the accuracy of security verification and lead to more advanced applications such as automatic modification of protocols.

We address the aforementioned two problems.
First, we propose a method to automatically generate training data, i.e., pairs of cryptographic protocols and security labels.
Specifically, we propose a method that automatically generates a random cryptographic protocol, converts the protocol to the input format of a formal verification tool, and obtains a security label.
This enables us to prepare arbitrarily large training datasets without manually collecting protocol data from the literature.
Moreover, we propose a method to construct a model that has both a sequence structure and a tree structure as a model that can handle cryptographic protocols (Figure~\ref{fig:protocol_feature_ours}).
This is based on the idea that a cryptographic protocol is a sequence of messages, i.e., has a series structure, and each message has a syntax tree structure with encryption operations, etc. as nodes.
The proposed model is suitable for capturing a hierarchical structure of protocols and thus should verify the security of protocols more correctly than the previous methods.
Furthermore, we show that the processing time of the model depends linearly on the size of protocols, which is desirable for practical use.
In summary, we propose a novel security verification framework of cryptographic protocols based on machine learning that provides practically fast, decidable, and accurate verification.

As a feasibility test, we implemented the proposed method and verified the confidentiality of the key exchange protocol using practical protocols. 
In the results, we achieved a 79.5\% verification accuracy.

The contributions of this paper are summarized as follows.
To construct a protocol verifier that is practically fast, decidable, and accurate, we do the following:
\begin{enumerate}
    \item We propose a method to automatically generate a large training dataset to achieve a highly accurate security verifier using machine learning for cryptographic protocols without manually collecting protocol data from the literature.
    \item We propose a model having both series and tree structures that is suitable for handling cryptographic protocols. The proposed model further improves the accuracy of verification.
    \item We validate our framework for security verification of cryptographic protocols based on the method in 1 and the model in 2 above by feasibility tests. The proposed method achieves 79.5 percent accuracy in verification of practical protocols.
\end{enumerate}

The paper is organized as follows: Section~\ref{sec:related} describes the related works, and Section~\ref{sec:preliminaries} briefly introduces cryptographic protocols, their formal verification techniques, and neural networks.
In Section~\ref{sec:proposed}, we introduce the proposed method.
Section~\ref{sec:experiment} describes the implementation and experiments, and Section~\ref{sec:conclusion} summarizes this paper.

\section{Related Work} 
\label{sec:related}

Several methods have been proposed to verify cryptographic protocols on the basis of machine learning~\cite{ma2020machine,zahednejad2020novel}.
Our work is novel in that it tackles two problems for constructing an accurate verification model: accessing numerous training data and exploiting detailed hierarchical features of protocols for learning.
Ma et al.~\cite{ma2020machine} were the first to construct an automatic verification framework for authentication and key exchange protocols using machine learning.
They proposed a method to transform protocols into a vector of fixed length (Figure~\ref{fig:protocol_feature_example}), created a training dataset from 500 protocols that had already been verified for security, and built a model to identify security using machine learning methods such as support vector machine (SVM).
Zahednejad et al.~\cite{zahednejad2020novel} 
proposed an improvement on Ma et al.'s transformation of protocols.
They also formulated the security verification as a multi-label classification instead of a multi-class classification and prepared about 1,000 training data, i.e., protocols with security labels to improve the performance of the classification model.
Both of these methods collect training data manually and ignore the hierarchical structure of messages in protocols.

Our approach uses machine learning to reduce the computational complexity of protocol verification.
Formal verification of cryptographic protocols has long been studied in terms of its decidability and computational complexity~\cite{durgin2004multiset,tiplea2005decidability,rusinowitch2003protocol,amadio2003symbolic,comon2005tree,abadi2006deciding,chretien2015decidability,ciobacua2009computing,ciobacua2012computing,conchinha2010efficient,4529462}, in particular to complete the computation in practical time by restricting the class of protocols and the adversary.
Although the verification is generally not decidable, some subclasses are.
However, they all restrict the modeling of the protocol or the ability of the adversary.
For example, verification is decidable when the number of sessions is bounded or when the adversary is passive, i.e., can eavesdrop on messages but cannot modify or retransmit them~\cite{4529462,tiplea2005decidability}.
Durgin et al.~\cite{durgin2004multiset} analyzed the complexity of verification under various restrictions and showed that several subclasses of protocols could be decidable.
DEEPSEC~\cite{8418623} provides a decision procedure and its tools for a limited number of sessions.
Our approach does not require these restrictions on the class of protocols and the adversary.

\section{Preliminaries}\label{sec:preliminaries} 
\subsection{Cryptographic Protocols}
Cryptographic protocols ensure the confidentiality and integrity of data flowing through a communication channel by using cryptographic primitives such as public-key cryptography and digital signatures.
Key exchange protocols, authentication protocols, and electronic voting protocols are examples of cryptographic protocols.
Here, we explain several concepts of cryptographic protocols, using key exchange protocols as an example.
A \emph{key exchange protocol} is for sharing a session key, which is a common secret key, with a communicating party over an unreliable communication channel such as the Internet~\cite{boyd2019protocols}.
The executor of a protocol is called a \emph{party}, and one action of executing the party's protocol (i.e., from the start of message exchange to the establishment of the session key) is called a \emph{session}.
We assume that there are two parties: the \emph{initiator} (denoted by $I$), who starts the session, and the \emph{responder} (denoted by $R$), who is the communication partner of the initiator.
Each party is assigned a unique identifier called a \emph{party ID}, a \emph{long-term secret key} (denoted by $\lsk$), and a \emph{public key} (denoted by $\pk$) corresponding to the long-term secret key, and each session is assigned a unique identifier called a session ID.
In addition, ephemeral secret information (e.g., random numbers generated in the session) used by each party in the protocol is called an \emph{ephemeral secret key} (denoted by $\esk$).
There are two types of key exchange protocols: transport, in which one party generates a session key and sends it to the other party, and establishment, in which both parties generate a session key using the information they exchange.

Key exchange protocols have two main security requirements: session key confidentiality and authentication.
Session key confidentiality means that an adversary cannot obtain all or part of the session key.
Authentication assures the correctness of the communicating party and prevents an adversary from impersonating the communicating party.

\subsection{Formal Verification of Cryptographic Protocols} 
In formal verification of cryptographic protocols, we describe the protocol specification and security model (i.e., adversary's model and security requirements) in a formal language and mathematically analyze whether the protocol specification satisfies the security model.

There exist formal verification tools that automatically obtain verification results, such as ProVerif~\cite{blanchet2013automatic,proverifweb}, Scyther~\cite{cremers2008scyther,scytherweb}, Tamarin prover~\cite{tamarinweb,meier2013tamarin}, and Verifpal~\cite{verifpalweb,kobeissi2020verifpal}.
These tools have successfully found vulnerabilities and ensured the security of various cryptographic protocols~\cite{basin2018formal,BhargavanBlanchetKobeissiSP2017,cremers2017comprehensive,cremers2016automated,whitefield2017formal,8970242,9170999}.
For example, formal verification tools have been used to verify TLS 1.3~\cite{cremers2016automated,cremers2017comprehensive,BhargavanBlanchetKobeissiSP2017,bhargavan2022symbolic} and 5G authentication protocols~\cite{basin2018formal,cispa2758,8970242,8923409}.

Formal verification, specifically symbolic formal verification techniques, provides security verification against the Dolev-Yao model~\cite{dolev1983security} and symbolically models protocol execution, and the adversary's behavior against it.
In the Dolev-Yao model, key information and protocol messages flowing in the communication channel are assumed as terms rather than strings.
Furthermore, cryptographic primitives such as public-key cryptosystems are formalized as idealized ones in which ciphertext can only be decrypted when the decryption key is possessed (i.e., no plaintext information can be obtained unless the decryption key is possessed).
The adversary is then modeled as an active adversary who can eavesdrop, tamper, delete, and retransmit any message that flows through the communication channel.

Formal verification of the Dolev-Yao model has been shown theoretically to be undecidable~\cite{durgin2004multiset,tiplea2005decidability}.
Therefore, verifiers need to manually add lemmas, restrict network configuration and adversary capabilities, etc., when verifying protocols.
However, this is a practical issue because it requires expert knowledge.

\subsection{Deep Learning Models}\label{subsec:deep_learning}
In this section, we briefly explain classification models based on neural networks and deep learning~\cite{Goodfellow-et-al-2016}.
Multi-class classification is the task of outputting an associated label for input data, and the classification model estimates the probability distribution $p(y\mid x)$ of label $y$ for input $x$.

A model structure of a neural network is generally defined by the repetition and combination of matrix products and nonlinear functions.
The most basic model taking a fixed-length vector as an input is known as \emph{multi-layer perceptron} (MLP).
Vectors computed in the deep learning model are called \emph{latent vectors} of the input data and considered to represent complex features of the data.
Depending on the structure and prior knowledge of the input data, a suitable model structure should be chosen.
Such a dependence is usually called an inductive bias.
For example, for image classification, a convolutional neural network is the most standard network structure.
In this section, we introduce two model structures that involve sequence and tree structures, respectively.

A \emph{recurrent neural network} (RNN) takes series data as input.
Let $(x_t)_{t=1}^T$ be the input series data, where $x_t \in \mathbb{R}^d$ is a real vector representing the data at time $t$.
Let $n$ be the dimension of the latent vector, and the latent vector $h_t \in \mathbb{R}^n$ (also called the state) of RNN is updated at each time $t=1,2,\cdots,T$ as follows:
\begin{align}
    h_t = \tanh(W x_t + U h_{t-1} + b),
\end{align}
where weight matrices $W \in \mathbb{R}^{n\times d}, U \in \mathbb{R}^{n\times n}$ and bias $b\in \mathbb{R}^n$ are the parameters of the model.
$\tanh$ is a hyperbolic tangent function, acting on each vector component.
The initial state $h_0$ is typically defined as the zero vector.
In this way, the latent vector $h_t$ of RNN represents complex temporal dependencies of the series data $(x_t)_{t=1}^T$.
Training of RNN is typically unstable, and thus a more sophisticated model called a \emph{long short-term memory} (LSTM)~\cite{lstm} is commonly used.
LSTM is a variant of RNN with another state $c_t$ called a memory cell, equipped with a \emph{gating mechanism} to control information flow by utilizing a gate function taking a value in the interval $[0,1]$.
The states in LSTM are updated through gate functions:
\begin{align}
    c_t &= f_t \odot c_{t-1} + i_t \odot \tanh(W x_t + U h_{t-1} + b), \\
    h_t &= o_t \odot \tanh(c_t),
\end{align}
where $f_t, i_t, o_t \in [0,1]^n$ are functions of the state $h_{t-1}$ and input $x_t$ called a forget, input, and output gate, respectively, and $\odot$ denotes the element-wise product of vectors.

Neural networks that process tree-structured data are constructed by extending neural networks that process serial data.
Child-Sum Tree-LSTM~\cite{tai-etal-2015-improved} (which we simply call \emph{Tree-LSTM}) is an extension of LSTM to a tree structure.
In the same way as the LSTM has a state $h_t$ and a memory cell $c_t$ at each time step $t$, each node $j$ is associated with a state $h_j$ and a memory cell $c_j$ in the Tree-LSTM that are computed on the basis of those of its child nodes.
Let $C(j)$ be the set of child nodes of node $j$. 
The state update rule for the latent vector $h_j\in \mathbb R^n$ of node $j$ and the memory cell $c_j\in \mathbb R^n$ is as follows:
\begin{align}
    \tilde h_j &= \sum_{k\in C(j)} h_k,  \\
    c_j        &= i_j \odot u_j + \sum_{k\in C(j)} f_{jk}\odot c_k, \\
    u_j &= \tanh(W_c x_j + U_c \tilde h_j + b_c), \label{eq:treelstm} \\
    h_j        &= o_j \odot \tanh(c_j),
\end{align}
where $x_j$ is an input vector of node $j$.
The vectors $i_j, f_{jk}, o_j \in [0,1]^n$ denote input, forget, and output gates computed from the states $h_j$ of the child node and the input $x_j$, respectively, controlling the information flow through the node.

Given a latent vector $h \in \mathbb R^n$ of the input $x$, the distribution of labels $p(y|x)$ is estimated by linearly transforming $h$ into a vector of dimensions of the number of label types and applying the softmax function
\begin{align}
    \mathrm{softmax} (x)_i = \frac{e^{x_i}}{\sum_j e^{x_j}}.
\end{align}
The prediction of the model is defined as the label for which this probability takes the maximum value.

In the training phase, the model parameters are optimized with respect to the loss function
\begin{align}
    L = \frac{1}{N} \sum_{i=1}^N -\log p(y_i|x_i)
\end{align}
on the basis of a gradient method, where $p(y_i | x_i)$ is the output distribution of the model.
The loss function is averaged over the training data $\mathcal D = \{(x_i,y_i)\}_{i=1}^N$ 
that consists of a pair of an input data $x_i$ and the label $y_i$ of $x_i$.

\section{Proposed Method}\label{sec:proposed} 
In this section, we propose a security verification framework of cryptographic protocols using machine learning.
Our framework has three processes.

\begin{figure}[t!]
\centering
\subfloat[Generate datasets using formal verification tool.]{
\includegraphics[width=0.36\paperwidth]{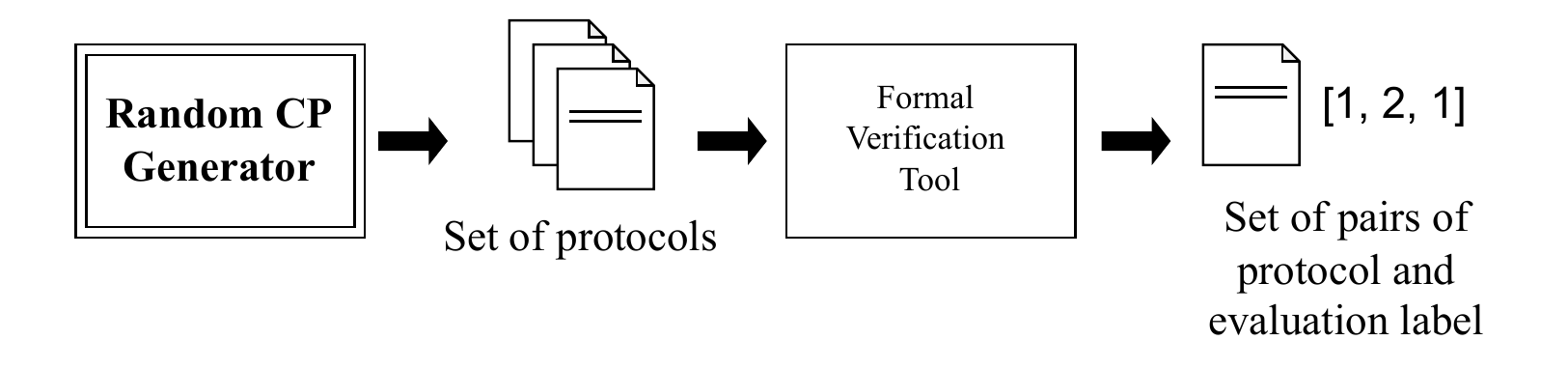}
\label{fig:framework1}}
\vspace{20pt}
\\
\subfloat[Generate verifier using machine learning.]{
\includegraphics[width=0.36\paperwidth]{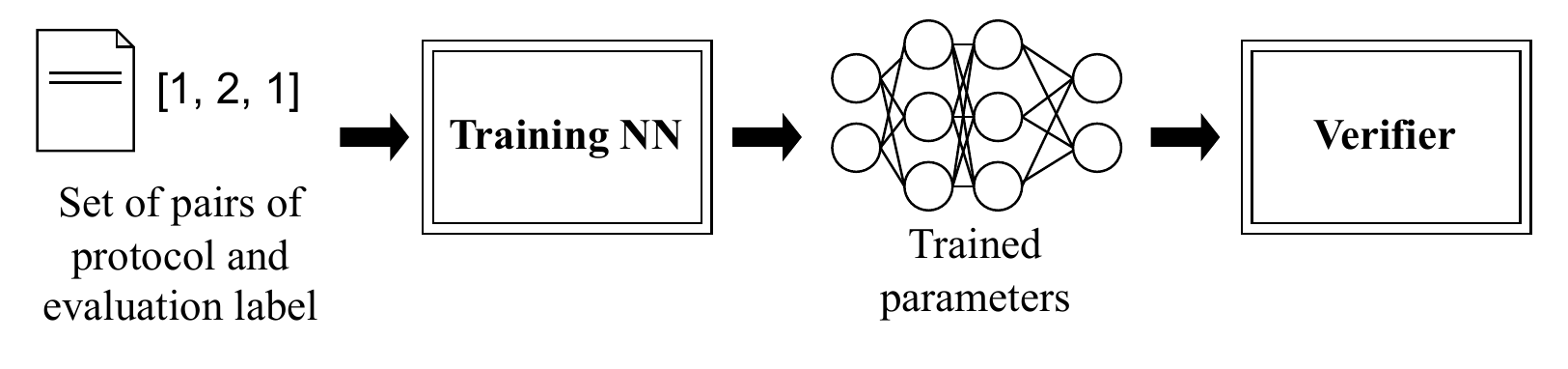}
\label{fig:framework2}}
\vspace{10pt}
\\
\subfloat[Security verification of cryptographic protocols using a verifier with trained parameters.]{
\includegraphics[width=0.36\paperwidth]{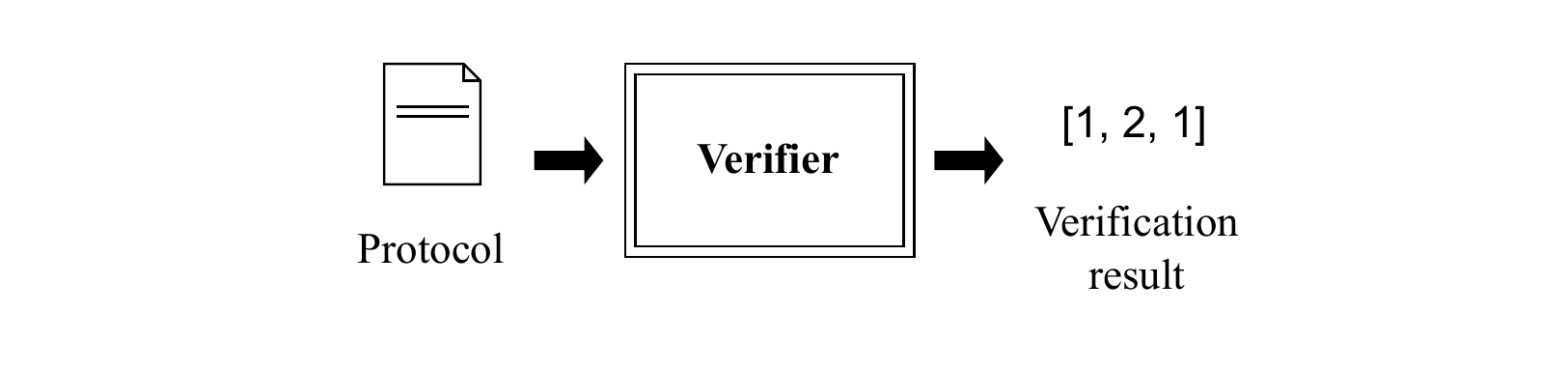}
\label{fig:framework3}}

\caption{Overview of our proposed framework.}
\label{fig:framework}
\end{figure}

The first is generating a dataset for training a machine learning model using formal verification (Figure~\ref{fig:framework1}).
This dataset is a set of pairs of a cryptographic protocol and its security evaluation label.
To generate the dataset, we construct a random automatic cryptographic protocol generator, see Section~\ref{subsec:dataset_generate}.
Generated protocols are input to the formal verification tool and assigned with security evaluation labels.
This process is fully automated and can generate over tens of thousands of training data.

The second process is to construct the protocol verifier using machine learning (Figure~\ref{fig:framework2}).
A machine-learning-based verifier is trained on the dataset constructed in the first process. 
We describe the construction of the model in Section~\ref{subsec:const_ml}.

The third process is the verification of cryptographic protocols using a trained verifier (Figure~\ref{fig:framework3}).
Users (e.g., the protocol designer) will verify the security of cryptographic protocols using the verifier constructed in the second process.
Using machine learning, users can verify cryptographic protocols in practical time, see Section~\ref{subsec:computational_time}.

To establish these processes, we utilize a tree representation of the abstract syntactic structure of messages in protocols.
In the first process, we utilize the tree structure of messages in protocols to automatically generate protocols and process them for verification.
Moreover, in the second process, we construct a neural network that reflects the tree and series structure of the cryptographic protocols.
In this way, protocols can be input into the machine learning model with no loss of structural features of the protocols.

\subsection{Representation of Cryptographic Protocols}\label{subsec:representation} 

In this section, we introduce representing protocols as sequences of abstract syntax tree for protocol generation and modeling.

\begin{figure}[t]
     \centering
\includegraphics[width=0.36\paperwidth]{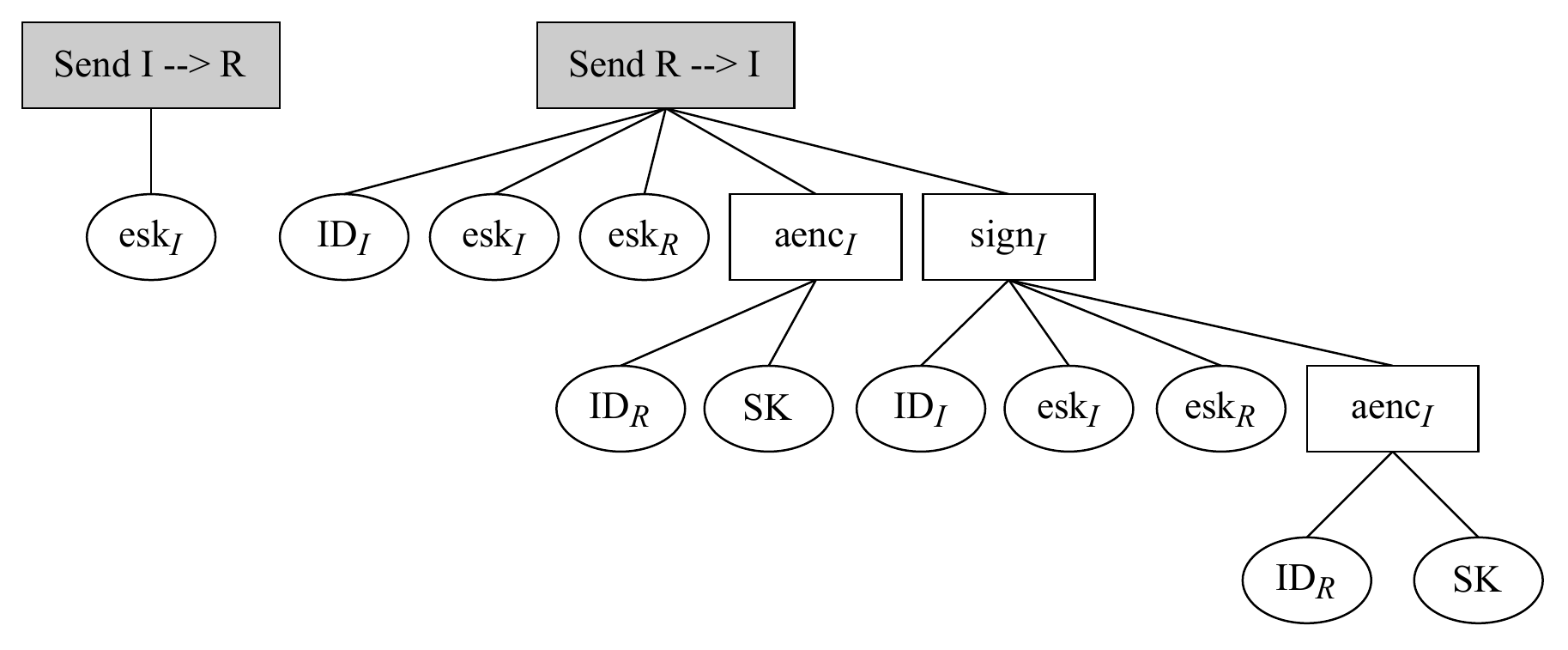}
     \caption{The representation of ISO/IEC 11770-3 Key Transport Mechanism 4~\cite{iso11770-3} using a syntax tree.
     The colored square nodes represent behavior nodes, the (white) square nodes represent function nodes, and the round nodes represent atomic nodes.}
     \label{fig:protocol_tree}
\end{figure}

Cryptographic protocols are represented by a sequence of messages exchanged between parties.
Each message of a cryptographic protocol is a string of letters in which operations such as hash functions and encryption functions are applied in sequence to message elements such as the party's ID and secret key.
Therefore, each message of a cryptographic protocol can be represented by a syntax tree that consists of a leaf node for message elements such as the party's ID and session key, an internal node for operations such as encryption, and a root node for message transmission.
Figure~\ref{fig:protocol_tree} shows an example of the representation of a cryptographic protocol using a syntax tree.
This figure represents the following authentication protocol (ISO/IEC 11770-3 Key Transport Mechanism 4~\cite{iso11770-3}).

\begin{align}\label{protocol:iso}
\begin{split}
  &I \rightarrow R: \esk_I \\
  &R \rightarrow I: \ID_I, \esk_I, \esk_R, \aenc_I (\ID_R, \SK), \\
  &\hspace{1.3cm} \sign_R (\ID_I, \esk_I, \esk_R, \aenc_I (\ID_R, \SK)).
\end{split}
\end{align}

Here, $\aenc_P$ means a public key encryption operation using party $P$'s public key, and $\sign_P$ means a signature generation operation using party $P$'s signature key (See Table~\ref{tab:structual_operation}).
The meanings of the elements of each message are shown in Table~\ref{tab:structual_element}.

We call an \emph{atomic node} an element of a message, such as the ID of a party or a generated random number. $\AN = \{a_1, \ldots, a_N\}$ denotes the set of atomic nodes.
We call a \emph{function node} such as encryption and digital signature creation operation nodes. $\FN = \{f_1, \ldots, f_M\}$ denotes the set of function nodes.
A party behavior, such as sending a message or accepting a session key, is called a \emph{behavior node}. $\BN = \{b_1, \ldots, b_L\}$ denotes the set of behavior nodes.
The elements of each node are defined in accordance with the protocol to be verified.
We define the elements we focus on in this paper in Section~\ref{subsec:target_protocol}.

\subsection{Generation of Dataset}\label{subsec:dataset_generate} 

To verify the security of a cryptographic protocol using machine learning, a large amount of training data must be prepared.
Since the protocols are few in number and must be obtained from scattered sources such as academic papers, they are not easy to manually collect. 
In this paper, we use a formal verification tool to generate such a dataset automatically.
First, we describe how to automatically generate the protocols in Section~\ref{subsec:autogen} and then how to attach evaluation labels to the protocols in Section~\ref{subsec:autolabel}.
Further improvement on the dataset generation is explained in Section~\ref{subsubsec:data_augmentation}.

\subsubsection{Automatic Generation of Protocols}\label{subsec:autogen}
To generate a training dataset, we need to generate a grammatically and functionally correct protocol automatically.
A protocol is represented as a sequence of messages composed of elements (e.g., party IDs and secret keys) and cryptographic primitives (e.g., encryption, signatures, etc.).
Therefore, we consider here a method to generate protocol messages in accordance with the protocol grammar.

\begin{algorithm}[t!]
\caption{\textsc{GenerateProtocol}}
\label{alg:protocolgen}
\begin{algorithmic}[1]
\Require Maximum number of message exchanges $m_{\max}$, maximum number of message elements $c_{\max}$, set of public atomic nodes $\AN_{\Pub}$
\Ensure Protocol $S$
\vspace{5pt}
    \State $\AN_I \coloneqq \AN_{\Pub} \cup \{ \esk_I, \lsk_I, \K\}$ \label{state:init_I}
    \State $\AN_R \coloneqq \AN_{\Pub} \cup \{ \esk_R, \lsk_R, \K\}$ \label{state:init_R}
    \State $m \xleftarrow{\$} \{ 1, \ldots, m_{\max}\}$ \label{state:sample_m_max}
    \For{$i = 1, \ldots, m$}
        \If{$i$ is odd} \label{state:set_root_node_start}
            \State $N_i \coloneqq \textsc{AddChild}(\_, \sendIR)$
            \State $P \leftarrow I$, $Q \leftarrow R$
        \ElsIf{$i$ is even}
            \State $N_i \coloneqq \textsc{AddChild}(\_, \sendRI)$
            \State $P \leftarrow R$, $Q \leftarrow I$
        \EndIf \label{state:set_root_node_end}

        \State \textsc{GenerateMessage}$(c_{\max}, \AN_{P}, P, N_{i})$ \label{state:generate_message}
        \State \textsc{UpdateKnowledge}$(\AN_{Q}, N_{i}, Q)$ \label{state:update_knowledge}
    \EndFor

    \State $S := [N_1, \ldots , N_{m_{\max}}]$
    \State \Return $S$
\end{algorithmic}
\end{algorithm}

\begin{algorithm}[t!]
\caption{\textsc{GenerateMessage}}
\label{alg:addchild}
\begin{algorithmic}[1]
\Require Maximum number of message elements $c_{\max}$, set of knowledge of $P$ $\AN_P$, party $P$, node $N$
\vspace{5pt}
\Function{GenerateMessage}{$c_{\max}$, $\AN_{P}$, $P$, $N$}
    \State $c \xleftarrow{\$} \{ 1, \ldots, c_{\max}\}$
    \For{$i = 1, \ldots, c$}
        \State $e \xleftarrow{\$} \AN_{P} \cup \FN$ \label{state:select_child_node_start}
        \State $N' = \textsc{AddChild} (N, e)$
        \If{$e \in \FN$}
            \State $\textsc{GenerateMessage}(c_{\max}, \AN_{P}, P, N')$
        \EndIf \label{state:select_child_node_end}
        
    \EndFor
\EndFunction
\end{algorithmic}
\end{algorithm}

We show pseudocodes of an algorithm for generating protocols in Algorithm~\ref{alg:protocolgen} and~\ref{alg:addchild}.
The outline is as follows:
\begin{enumerate}
    \item Determine the maximum number of message exchanges $m_{max}$ and the maximum number of message elements $c_{max}$.
    \item Randomly determine the number of message exchanges from $1, \ldots, m_{max}$ (line~\ref{state:sample_m_max} in Algorithm~\ref{alg:protocolgen}).
    \item Set the root node of each message alternately to the operation of sending messages from $I$ to $R$ and from $R$ to $I$ (lines~\ref{state:set_root_node_start} to \ref{state:set_root_node_end} in Algorithm~\ref{alg:protocolgen}).
    \item 
    Randomly select a child node element from the selectable atomic nodes or function nodes and repeat the process until the message is complete. (lines~\ref{state:select_child_node_start} to \ref{state:select_child_node_end} in Algorithm~\ref{alg:addchild}).
\end{enumerate}

Each party is allowed to use only the information it knows in messages it sends.
This set of information is called the knowledge set, consisting of atomic nodes and generated messages such as ciphertexts.
In step 4 above, each party selects elements from its own knowledge set.
When the party receives messages, the knowledge set is updated on the basis of the received messages.  
In addition to the received messages themselves, 
information that can be computed using the information in the knowledge set is added to the knowledge set.
For example, if the knowledge set contains a ciphertext and its corresponding secret key, then the plaintext of the ciphertext is added to the knowledge set.
$\AN_{\Pub}, \AN_I, \AN_R \subseteq \AN$ in the algorithms represent the knowledge sets of parties.
$\AN_{\Pub}$ is a set of public information from atomic nodes, including party's ID, public keys, and timestamps.
$\AN_I$ and $\AN_R$ are the sets of information known by parties $I$ and $R$, respectively.
These sets contain unique information for each party such as secret keys.
The symbol $\xleftarrow{\$}$ means uniform random choice from a set.

The algorithm \textsc{GenerateProtocol} takes as input the maximum number of messages $m_{\max}$, the maximum number of child nodes $c_{\max}$, and the set $\AN_{\Pub}$, and $\FN$, and outputs the protocol $N$.
First, the knowledge set $\AN_I, \AN_R$ is set as $\AN_{\Pub} \cup \{ \esk_P, \lsk_P, \K\}$ ($P \in \{I, R\}$) (lines~\ref{state:init_I} and \ref{state:init_R}).
Then, the number of messages $m$ is randomly determined between $1$ to $m_{\max}$ (line~\ref{state:sample_m_max}).
Each message is represented by a syntax tree. 
The root node of each message is set alternately by $\mathrm{sendIR}$ and $\mathrm{sendRI}$, which means send a message from $I$ to $R$ (from $R$ to $I$) (lines~\ref{state:set_root_node_start} to \ref{state:set_root_node_end}).
Here, \textsc{AddChild}$(N, e)$ is a function that adds a child node labeled with $e \in \AN_P \cup \FN$ to the node $N$ and returns the child node.
After that, a message is constructed by adding a child node to the root node (line~\ref{state:generate_message}).
\textsc{GenerateMessage} function (see Algorithm~\ref{alg:addchild}) generates protocol messages in accordance with the protocol grammar.
Details are described below.
Then, the knowledge set of the communication partner is updated in accordance with the received messages (line~\ref{state:update_knowledge}).
\textsc{UpdateKnowledge} function updates the knowledge set of party $Q$ with the information of message $N$ from $\AN_Q$.
This can be implemented straightforwardly by reading the generated message $N_i$ in the reverse order, that is, from the root to the leaf nodes, and collecting elements unknown to the party $Q$.
The above process is repeated for the number of message sentences to generate $m$ messages.
Finally, they are concatenated and the protocol is output as $N := [N_1, \ldots , N_{m_{\max}}]$.

The recursive function \textsc{GenerateMessage} takes the maximum number of message elements $c_{\max}$, the knowledge set of party P $\AN_P$, and node $N$, and adds $c$ child nodes to node $N$. 
$c$ is randomly selected from $\{1, \ldots ,c_{\max}\}$.

By generating protocols using the above process, we can automatically generate protocols.
We show an example of 
protocol generation in the following.

\begin{example*}

We assume $m_{max} = 3$ and $c_{max} = 5$.
Next, the number of message exchanges is randomly determined. Here, we assume that $2 \in \{1, 2, 3 \}$ is chosen as the number of rounds.
\begin{align*}
I \rightarrow R &: \ldots \\
R \rightarrow I &: \ldots \\
\end{align*}
Next, the number of elements in the first message is randomly determined. Here, we assume that $3 \in \{1, \ldots , 5 \}$ is chosen.
\begin{align*}
I \rightarrow R &: \_ \;,  \; \_ \; , \; \_ \\
R \rightarrow I &: \ldots \\
\end{align*}
Next, the parent node of the first element, i.e., the outermost operation, is randomly determined.
Here, we assume that public key encryption $\aenc \in \FN \cup \AN_I$ is chosen.
\begin{align*}
I \rightarrow R &: \aenc (\ldots, \pk_R) \;,  \; \_ \; , \; \_ \\
R \rightarrow I &: \ldots \\
\end{align*}
Then, the number of message elements to be subjected to public key encryption is randomly determined. Here, we assume that $1 \in \{1, \ldots , 5 \}$ is chosen at random.
\begin{align*}
I \rightarrow R &: \aenc (\_ \;, \pk_R) \;,  \; \_ \; , \; \_ \\
R \rightarrow I &: \ldots \\
\end{align*}
Next, the message to be subjected to public key encryption is randomly determined from the information known to the initiator.
Note that since this is the first message, the initiator only knows the public information and its own information.
Here, we assume that $\ID_R \in \FN \cup \AN_I$ is chosen.
\begin{align*}
I \rightarrow R &: \aenc (\ID_R, \pk_R) \;,  \; \_ \; , \; \_ \\
R \rightarrow I &: \ldots \\
\end{align*}
The other two elements of the first message are determined in the same way.
Here we assume that the following selections are made.
\begin{align*}
I \rightarrow R &: \aenc (\ID_R, \pk_R) \;, \hash (\senc (\esk_I, \K)) \; , \esk_I \\
R \rightarrow I &: \ldots \\
\end{align*}
The first message (from I to R) is now complete.
Then, the knowledge set of R is updated on the basis of the first message, and the next message (from R to I) is created in the same way.
\begin{align*}
I \rightarrow R &: \aenc (\ID_R, \pk_R) \;, \hash (\senc (\esk_I, \K)) \; , \esk_I \\
R \rightarrow I &: \hash (\esk_I), \ID_I \\
\end{align*}
When generating an establishment-type key exchange protocol, determining what kind of key to accept as a session key after sending and receiving (exchanging) messages is necessary.
The above algorithms can easily be extended to such cases.
Specifically, the session key is chosen randomly from the intersection $\AN_I \cap \AN_R$ of the knowledge set of the parties.
The following is an example:
\begin{align*}
I \rightarrow R &: \aenc (\ID_R, \pk_R) \;, \hash (\senc (\esk_I, \K)) \; , \esk_I \\
R \rightarrow I &: \hash (\esk_I), \ID_I \\
\mathrm{acceptI} &: \hash (\esk_I, \pk_R) \\
\mathrm{acceptR} &: \hash (\esk_I, \pk_R) \\
\end{align*}
\end{example*}

\subsubsection{Automatic Evaluation Labeling of Protocols}\label{subsec:autolabel} 
For a set of random protocols generated by the method described in the previous section, the labels for security evaluation are obtained by converting them to the input format of a formal verification tool and inputting them to the tool.
The method of conversion to the input format of the formal verification tool strongly depends on the tool. In most cases, a protocol tree structure can be used for successful conversion.
For example, the ISO/IEC 11770-3 Key Transport Mechanism 4 (see Section~\ref{subsec:representation}) in the input format of the formal verification tool ProVerif is shown in Code~\ref{code:iso}.

\begin{lstlisting}[caption=ProVerif Code of Protocol~\ref{protocol:iso},label=code:iso]
(* Initiator *)
let Initiator(lsk_I:bitstring, pk_I:bitstring, pk_R:bitstring, K:bitstring) =
  event protocol_start_I(ID_R);
  new esk_I1:bitstring;
  new esk_I2:bitstring;
  new T_I:bitstring;

  out (c, T_I);
  in (c, T_R:bitstring);

  (* AUTOMATIC_I *)
  out(c,(esk_I1));
  in(c, (=ID_I,=esk_I1,esk_R1:bitstring,m5:bitstring,m6:bitstring));
  let (=ID_R,SK:bitstring) = adec(m5,lsk_I) in
  if verif(m6, (ID_I,esk_I1,esk_R1,aenc((ID_R,SK),pk_I)),pk_R)=true then
  event acceptI(SK);
  out(c, senc(message, SK)).

(* Responder *)
let Responder(lsk_R:bitstring, pk_R:bitstring, pk_I:bitstring, K:bitstring) =
  event protocol_start_R(ID_I);
  new esk_R1:bitstring;
  new esk_R2:bitstring;
  new SK:bitstring;
  new T_R:bitstring;

  in (c, T_I:bitstring);
  out (c, T_R);

  (* AUTOMATIC_R *)
  in(c, (esk_I1:bitstring));
  out(c,(ID_I,esk_I1,esk_R1,aenc((ID_R,SK),pk_I),sign((ID_I,esk_I1,esk_R1,aenc((ID_R,SK),pk_I)),lsk_R)));
  event acceptR(SK).
\end{lstlisting}
Lines 1 to 19 represent the initiator's behavior.
Lines 11 to 15 are the code that is automatically converted from the protocol, which includes sending (line 12) and receiving (line 13) messages, decrypting the ciphertext (line 14), and verifying the signature (line 15).
Lines 22 to 38 represent the responder behavior, and lines 33 to 35 represent the part that is automatically converted from the protocol.
For details of the ProVerif grammar, please refer to the official manual~\cite{blanchet2020proverif}.
This process allows automatic collection of protocol-security label pairs without having to manually collect protocols from papers.

\subsubsection{Data Augmentation}\label{subsubsec:data_augmentation}

Automatic protocol generation in the above method results in a high probability of insecure and unnatural protocols.
If those protocols are used as they are for learning, 
learning verification tends to fail since the data distributions of generated protocols and practical ones are much different.
Therefore, we further utilize suitable augmentation of protocol data to improve training.

Machine learning models are good at capturing patterns in training data.
To train our models to learn patterns in vulnerable protocols, we change parts of secure protocols to make them insecure 
and use them as training data.
Here, we convert a secure protocol into a similar insecure one in the following way:
\begin{itemize}
    \item Sending the session key or secret key over the communication channel.
    \item Using keys known to the adversary for encryption.
    \item Creating a session key from information known to the adversary.
\end{itemize}

For example, modifying protocol~\ref{protocol:iso} in the first way would result in the following:

\begin{align*}
  &I \rightarrow R: \esk_I \\
  &R \rightarrow I: \ID_I, \esk_I, \esk_R, \aenc_I (\ID_R, \SK), \\
  &\hspace{1.3cm} \sign_R (\ID_I, \esk_I, \esk_R, \aenc_I (\ID_R, \SK)), \SK.
\end{align*}

Thus, by creating pairs of similar secure/insecure protocols
 the model can learn the patterns that make a protocol insecure, which is expected to improve the accuracy of verification.

\subsection{Construction of Machine Learning Model}\label{subsec:const_ml} 
Common machine learning models such as SVM and MLP
take fixed-length vectors as input.
Therefore, cryptographic protocols must be converted to fixed-length vectors in some way to be taken as inputs~\cite{ma2020machine,zahednejad2020novel}.
Existing transformation methods such as TLM~\cite{ma2020machine} ignore the grammatical structure of the message and cause the loss of structural information of the cryptographic protocol, thus reducing the accuracy of the security verification.
For example, in existing methods, the protocols
\[
I \rightarrow R: \mathrm{sign}_I (\mathrm{aenc}_R (\ID_I, \SK))
\]
and 
\[
I \rightarrow R: \mathrm{aenc}_R ( \mathrm{sign}_I (\ID_I, \SK))
\]
are converted to the same vector.
That is, the converted vectors cannot be distinguished between protocols.
The elements of each message are shown in Table~\ref{tab:structual_element}.
In this section, we propose a new machine learning model that can accept cryptographic protocols directly by constructing a neural network that reflects the data structure of the protocols.

\begin{figure}[t]
     \centering
\includegraphics[width=0.36\paperwidth]{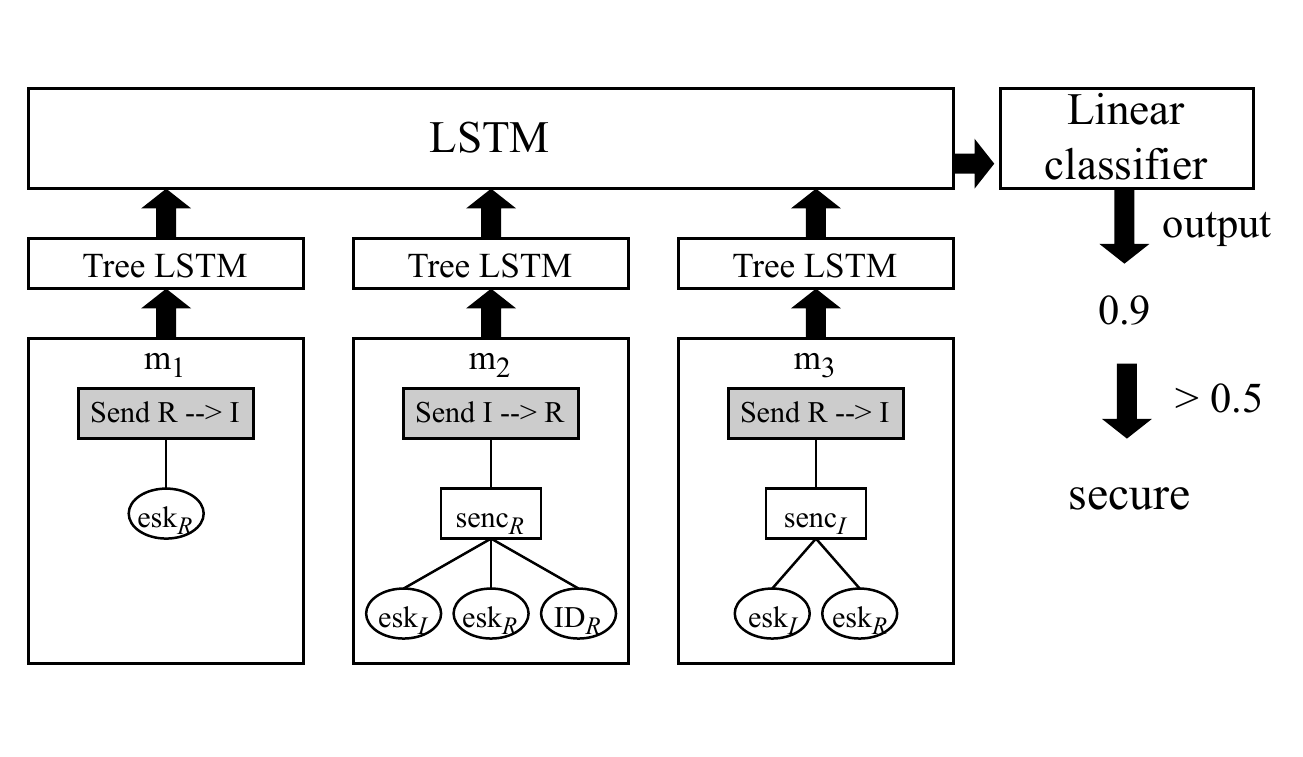}
     \caption{A neural network for processing protocols. Each message is processed by a Tree-LSTM with a tree structure, and its output is processed by the LSTM. Then, through the linear classification layer, which is the output layer, the prediction of the security label is output as a probability.}
     \label{fig:protocol_net}
\end{figure}

As described in Section~\ref{subsec:representation}, each message of a cryptographic protocol can be represented by a syntax tree.
Therefore, we construct a new neural network that processes cryptographic protocols by combining neural networks that deal with the sequence and tree structures (see Figures~\ref{fig:protocol_feature_ours} and~\ref{fig:protocol_net}).
That is,
the proposed model processes a protocol in two stages: 
it first takes each message with the tree structure as an input and then processes the latent vector of each message in sequence.
To this end, we utilize the Tree-LSTM for the first stage and LSTM for the second stage.
Since each node in the tree should be associated with a vector, i.e., $x_j$ in Equation \ref{eq:treelstm}, to be an input to the Tree-LSTM model, we assign a random vector sampled from normal distribution to each behavior, function, and atomic node.
After every message in a protocol is processed in the Tree-LSTM, the sequence of obtained outputs is passed as the input to the LSTM.
Using the output of the LSTM, the protocol is classified as secure or vulnerable through a linear output layer.
The whole model is trained in a usual way by minimizing the cross-entropy loss with gradient methods.

\subsection{Computational Complexity of Verification}\label{subsec:computational_time}

After completion of training, the model is used for verification.
The main advantage of our framework is that the verification can always terminate in practical time.
Specifically, we show that our model processes a protocol with a linear order of time with respect to the size of the protocol.
Here, the size of the protocol is defined as the total number of nodes in the messages.

First, the Tree-LSTM in our model processes each message in a protocol, reading each node in the message exactly once along the tree structure of the message.
At each node, the Tree-LSTM requires a constant amount of computational time under the state update rule~(\ref{eq:treelstm}).
Thus, the Tree-LSTM processes a message in a protocol with a linear order of time with respect to the number of nodes in the message.
After that, the LSTM in our model processes the latent feature of each message returned by the Tree-LSTM along the sequence structure of the protocol.
The computational time of the LSTM is proportional to the sequence length, i.e., the number of messages in a protocol, which is bounded by the total number of nodes in the messages.
Thus, the total computational time of our model has a computational complexity of the linear order with respect to the size of the protocol.

\section{Experiments}\label{sec:experiment} 
To validate the effectiveness of the proposed method, we evaluate the performance of the security verification of cryptographic protocols using the training data generated by the method in Section~\ref{subsec:dataset_generate} and the machine learning model constructed by the method in Section~\ref{subsec:const_ml}.

\subsection{Target Protocols and Security Requirement}\label{subsec:target_protocol}
Our target protocols and security requirements to be verified in this study are the two-party key exchange protocol and the session key confidentiality requirements.
This is the most standard of all cryptographic protocols.

We cover both transport and establishment type key exchange protocols.
We set the maximum number $m_{\max}$ of messages to 5 and the maximum number of child nodes $c_{\max}$ to 3.
The following describes the elements that compose the messages and party behavior of the target protocol.

\subsubsection{Protocol Message Components}
The components of protocol messages and their notation are shown in Table~\ref{tab:structual_element}.
Here $P$ in the table represents initiator $I$ or responder $R$, i.e., $P \in \{I, R\}$.
An ephemeral secret key $\esk$ means session-specific secret information such as random numbers, and a static secret key means a secret key used for decryption of public key encryption, which is known only to the corresponding party, i.e., $\esk_P$ and $\lsk_P$ are known only to party $P$.
The party ID $\ID_P$, the public key $\pk_P$ used to encrypt the public key cryptography, and the timestamp $T$ are public information, and are known to parties $I$ and $R$ as well as the adversary.
The pre-shared secret key $\K$ is the secret key used to encrypt and decrypt the symmetric key cryptography, and is known only to parties $I$ and $R$.
The session key $\SK$ is shared between parties through message exchange.

\begin{table}[t]
\begin{center}\small
\caption{Protocol Message Components}
\begin{tabular}{p{5.2cm} p{2.3cm}} \toprule
	Component & Notation\\ \midrule
	ID of party $P$ & $\ID_P$ \\
	ephemeral secret key of party $P$ & $\esk_P$ \\
	static secret key of party $P$ & $\lsk_P$ \\
	public key of party $P$ & $\pk_P$ \\
	timestamp of party $P$ & $T_P$ \\
	pre-shared secret key of parties & $\K$ \\
	session key of party $P$ & $\SK_P$ \\
\bottomrule
\label{tab:structual_element}
\end{tabular}
\end{center}
\end{table}

\subsubsection{Protocol Message Structure}
The operations of the parties in composing the message and their notations are shown in Table~\ref{tab:structual_operation}.
Here $\Vec{m}$ in the table represents the sequence of messages, i.e., those generated from the components of the message on the basis of composition method.

We assume that the pre-shared secret key $\K$ is used as the symmetric key when generating the symmetric key ciphertext, the public key $\pk_Q$ of communicating party $Q$ is used as the public key when generating the public key ciphertext, and party $P$'s long-term private key $\lsk_P$ is used as the private key when generating the digital signature.

\begin{table}[t]
\begin{center}\small
\caption{Message Structure}
\begin{tabular}{p{5.2cm} p{2.3cm}} \toprule
	Operation & Notation\\ \midrule
	Symmetric key ciphertext of $\vec{m}$ using symmetric key $\K$ & $\mathrm{senc} (\vec{m}; \K)$ \\
	Public key ciphertext of $\vec{m}$ using public key $\pk$ & $\mathrm{aenc} (\vec{m}; \pk)$ \\
	Digital signature of $\vec{m}$ using static secret key $\lsk$ & $\mathrm{sign} (\vec{m}; \lsk)$ \\
	Hash value of $\vec{m}$ & $\mathrm{hash} (\vec{m})$\\
 	Exponent ($m_0$ to the power of $m_1$) & $\mathrm{exp} (m_0; m_1)$\\
\bottomrule
\label{tab:structual_operation}
\end{tabular}
\end{center}
\end{table}

\subsubsection{Party Behavior}
The party behaviors in the protocol and their notation are shown in Table~\ref{tab:party_behavior}.
Note that $m$ in the table denotes an arbitrary message.

As party operations, we consider the operations of public key encryption, symmetric key encryption, digital signature, and hash value calculation for some values (e.g., a message).
The value that is the target of the operation can take the concatenation of multiple values. 

\begin{table}[t]
\begin{center}\small
\caption{Party Behavior}
\begin{tabular}{p{5.2cm} p{2.3cm}} \toprule
	Behavior & Notation\\ \midrule
	Send $m$ from $I$ to $R$ & $\mathrm{sendIR} (m)$ \\
	Send $m$ from $R$ to $I$ & $\mathrm{sendRI} (m)$ \\
	$I$ accepts $m$ as session key & $\mathrm{acceptI} (m)$ \\
	$R$ accepts $m$ as session key & $\mathrm{acceptR} (m)$ \\
\bottomrule
\label{tab:party_behavior}
\end{tabular}
\end{center}
\end{table}

\subsection{Evaluation Metrics}

We evaluate verification accuracy of trained models with two types of test datasets.
One consists of randomly generated protocols (\emph{Random}) and the other consists of manually collected practical protocols (\emph{Practical}).
The former includes an equal number of secure and insecure protocols randomly generated by a method explained in Section~\ref{subsec:dataset_generate} (without augmentation).
The number of such protocols is 384.
The latter consists of protocols collected manually from a textbook~\cite{boyd2019protocols}.
There are 26 protocols subject to our verification setting in Section~\ref{subsec:target_protocol}.
The protocols used for evaluation and their security labels are listed in Table~\ref{tab:protocol_list}.
The numbers in the table represent the protocol numbers in the textbook~\cite{boyd2019protocols}.
Since most of them are secure, a model judging all protocols as secure achieves high accuracy on these data, which is not desirable.
Therefore, we make insecure protocols by manipulation similar to the augmentation in Section~\ref{subsubsec:data_augmentation} and add them to the test data, which increases the number of practical test data to 44.

\begin{table*}[t]
  \centering
  \caption{Protocols Used for Evaluation~\cite{boyd2019protocols}}\label{tab:protocol_list}
  \begin{tabular}{cccc}
    \hline
    Protocol & No. & Round & Attack found \\
    \hline
    STS protocol & 1.9 & 3 & no \\
    STS protocol modified to include identifiers & 1.10 & 3 & no \\
    Protocol ntor of Goldberg, Stebila and Ustaoglu & 1.15 & 2 & no \\
    Revised Andrew protocol of Burrows et al. & 3.14 & 4 & yes \\
    Boyd two-pass protocol & 3.16 & 2 & no \\
    ISO/IEC 11770-2 Key Establishment Mechanism 1 & 3.17 & 1 & no \\
    ISO/IEC 11770-2 Key Establishment Mechanism 2 & 3.18 & 1 & no \\
    ISO/IEC 11770-2 Key Establishment Mechanism 3 & 3.19 & 1 & no \\
    ISO/IEC 11770-2 Key Establishment Mechanism 4 & 3.20 & 2 & no \\
    ISO/IEC 11770-2 Key Establishment Mechanism 5 & 3.21 & 2 & no \\
    ISO/IEC 11770-2 Key Establishment Mechanism 6 & 3.22 & 3 & no \\
    ISO/IEC 11770-3 Key Transport Mechanism 1 & 4.11 & 1 & no \\
    ISO/IEC 11770-3 Key Transport Mechanism 2 & 4.12 & 1 & no \\
    ISO/IEC 11770-3 Key Transport Mechanism 3 & 4.13 & 1 & no \\
    Denning-Sacco public key protocol & 4.14 & 1 & no \\
    ISO/IEC 11770-3 Key Transport Mechanism 4 & 4.15 & 2 & no \\
    ISO/IEC 11770-3 Key Transport Mechanism 5 & 4.16 & 3 & no \\
    ISO/IEC 11770-3 Key Transport Mechanism 6 & 4.17 & 3 & no \\
    Helsinki protocol & 4.18 & 3 & yes \\
    Blake-Wilson-Menezes key transport protocol & 4.19 & 3 & no \\
    Needham-Schroeder public key protocol & 4.20 & 3 & yes \\
    Needham-Schroeder-Lowe protocol modeified by Basin et al. & 4.22 & 3 & no \\
    X.509 one-pass authentication & 4.24 & 1 & no \\
    X.509 two-pass authentication & 4.26 & 2 & no \\
    X.509 three-pass authentication & 4.27 & 3 & no \\
    Diffie-Hellman key agreement & 5.1 & 2 & yes \\
    \hline
  \end{tabular}
\end{table*}

\subsection{Training Setup}

For creating a training dataset,
we generate 
50,000 protocols following the method described in Section~\ref{subsec:autogen}. 
Then, we use ProVerif~\cite{blanchet2013automatic,proverifweb}, a formal verification tool of cryptographic protocols, to assign security evaluation labels to each generated protocol.
ProVerif does not ensure decidability and may require a large amount of verification time, depending on the protocol.
In this experiment, we only deal with protocols that can be verified within five seconds. 
After the labeling, we collect the protocols verified as secure and then generate insecure protocols by data augmentation explained in Section~\ref{subsubsec:data_augmentation}.
The resulting number of the training data is 16,371.

We train the model proposed in Section~\ref{subsec:const_ml} on the training dataset.
The number of dimensions of the latent vector of the LSTM and Tree-LSTM in the model is set to 128.
We use RMSprop with batch size 100 to optimize the model with 200 parameter updates.
The learning rate is set to 0.001.

As baselines, we apply the previous methods in \cite{ma2020machine,zahednejad2020novel} using the same generated training dataset.
We used TLM~\cite{ma2020machine} and the method of Zahednejad et al~\cite{zahednejad2020novel} for converting the training data to vectors of fixed size.
Since their conversion rules (e.g., which element corresponds to which scalar) are not fully explained in either the papers or the public codes\footnote{https://github.com/zahednejad/protocol-analysis-with-machinelearning} of~\cite{zahednejad2020novel}, we re-implemented these methods on our own.
For learning models, we adopted XGBoost and 3-layer MLP implementation from the public codes of \cite{zahednejad2020novel}.

\subsection{Results}

\begin{table}[t]
  \centering
  \caption{Accuracy of Verification on Test Sets}\label{tab:verification_result}
  \begin{tabular}{ccrr}
    \hline
    Conversion & Model & Random & Practical \\
    \hline
    TLM & MLP       & 71.6 & 52.3 \\
    TLM & Xgboost   & 72.3  & 54.6 \\
    \cite{zahednejad2020novel} & MLP     & 58.8 & 52.3 \\
    \cite{zahednejad2020novel} & Xgboost & 62.5 & 50.0 \\
    \textbf{N/A} & \textbf{Ours}     & \textbf{85.3} & \textbf{79.5} \\
    \hline
  \end{tabular}
\end{table}

We evaluate the performance of verification by a trained model on the test datasets.
The overall accuracy is shown in Table~\ref{tab:verification_result}.
Note that previous methods utilize conversion methods of translating protocols into vectors, 
whereas our proposed model does not require conversion since protocols are input into the model with its original tree and sequential structures.
The models with previous methods on modeling protocols~\cite{ma2020machine,zahednejad2020novel} verify randomly generated protocols correctly to some extent, though they almost completely fail to verify practical protocols.
TLM~\cite{ma2020machine} performs better than the method of Zahednejad et al.~\cite{zahednejad2020novel}.
This indicates that TLM provides more natural modeling of protocols on the basis of their structures than the method of Zahednejad et al.
Our proposed model performs best on both randomly generated protocols and practical protocols.
These results show that our model has a better inductive bias for dealing with cryptographic protocols.

\begin{figure}[t]
     \centering
     \includegraphics[width=0.36\paperwidth]{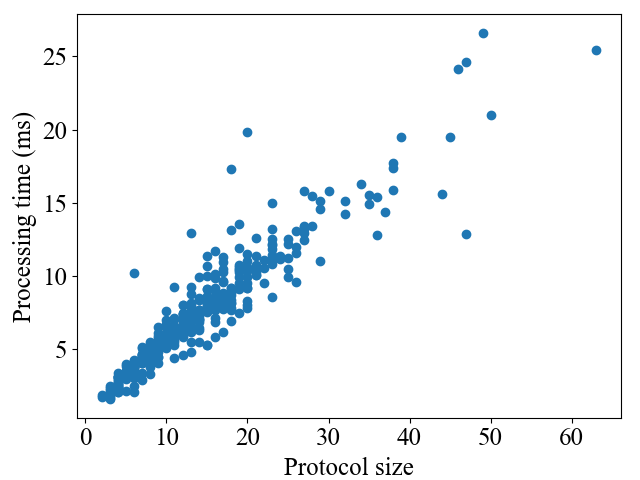}
     \caption{Processing time for verification with respect to protocol size.}
     \label{fig:processing_time}
\end{figure}

As explained in Section~\ref{subsec:computational_time}, the processing time grows linearly with respect to protocol size in theory,
where the protocol size is defined as the total number of nodes of syntax trees of messages in a protocol.
To confirm this property, we further measured processing time of our model on each protocol in the randomly generated test set.
The results in Figure~\ref{fig:processing_time} show that the processing time indeed grows linearly with respect to the protocol size.
Therefore, our proposed model can manage verification on large protocols.

\section{Conclusion}\label{sec:conclusion} 
In this paper, to construct a security verifier for cryptographic protocols with practical speed, we proposed a security verification framework for cryptographic protocols using machine learning.
To improve the accuracy of verification, we established a method to generate cryptographic protocols and add security labels to the generated protocols using a formal verification tool to automatically generate large amounts of training data for machine learning.
Furthermore, we constructed a model able to handle both tree and series structures that is suitable for processing cryptographic protocols.
We implemented the proposed method, conducted a feasibility test for practical protocols, and achieved an accuracy of 79.5\%.
These results indicate that the proposed verification method is more accurate than the previous method with a practical verification time.

Future issues include further improvement of verification accuracy.
To achieve this, we consider that a method will need to be established for automatically generating training data that more closely follows actual protocols.
Furthermore, future prospects for the application of machine learning to the field of cryptographic protocols include inferring the cause of protocol vulnerabilities, automatically fixing vulnerabilities, and automatically generating protocols that satisfy specified requirements.
We constructed a verifier in this study, but would like to link it to these studies in the future.

\bibliographystyle{plain}
\bibliography{main}

\end{document}